\documentclass[aps,twocolumn,showpacs,letterpaper,prl]{revtex4}

\usepackage{graphicx,amsmath}

\begin{document}

\def\pcco{Pr$_{1.85}$Ce$_{0.15}$CuO$_{4-y}$ }
\def\htc{high-$T_c$ }
\def\cu{$^{63}$Cu}
\def\cuf{$^{65}$Cu}
\def\lsco{Ln$_{2-x}$Sr$_x$CuO$_4$}
\def\invtone{$T_1^{-1}$}
\def\ttg{$T_{2G}$}
\def\ttr{$T_{2R}$}
\def\a{\textbf{a}}
\def\b{\textbf{b}}
\def\c{\textbf{c}}

\title{Inhomogeneous electronic structure probed by spin-echo 
experiments in the electron doped high-$T_c$ superconductor \pcco}

\author{F.~Zamborszky$^1$, G.~Wu$^1$, J.~Shinagawa$^1$, W.~Yu$^1$,
H.~Balci$^2$,  R.~L.~Greene$^2$, W.~G.~Clark$^1$, S.~E.~Brown$^1$}

\affiliation{$^1$ Department of Physics and Astronomy, UCLA, Los
Angeles, California 90095-1547}

\affiliation{$^2$ Department of Physics and Center of Superconductivity
Research, University of Maryland, College Park, Maryland 20742}

\date{\today}

\begin{abstract}
\cu\ nuclear magnetic resonance (NMR) spin-echo decay rate ($T_2^{-1}$)
measurements are reported for the normal and superconducting states of a
single crystal of \pcco (PCCO) in a magnetic field $B_0=9$~T over the
temperature range 2~K$<T<$200~K. The spin-echo decay rate is
temperature-dependent for $T<55$~K, and has a substantial dependence on
the radio frequency (rf) pulse parameters below $T\approx 25$~K. This
dependence indicates that $T_2^{-1}$ is strongly effected by a local
magnetic field distribution that can be modified by the rf pulses,
including ones that are not at the nuclear Larmor frequency. The
low-temperature results are consistent with the formation of a static
inhomogeneous electronic structure that couples to the rf fields of the
pulses.

\pacs{74.72.Jt,76.60.-k,74.25.Jb}

\end{abstract}

\maketitle

Understanding the normal phase of high-temperature cuprate
superconductors is widely recognized as essential to a successful theory
of the \htc problem. Important points include the origin and properties
of a pseudogap, the unusual temperature dependence of the in-plane
resistivity, and other unconventional properties. For this
reason, it is important to study the properties of the normal phase even 
in the low-temperature limit. For most materials, the very large
upper critical-fields $B_{c2}$ make this desirable condition
unattainable. $B_{c2}$ is significantly smaller in the electron-doped
cuprates. In particular, $B_{c2}\approx 10$~T for Pr$_{2-x}$Ce$_x$CuO$_{4-y}$ 
at optimal doping ($x\approx0.15$) when the field is applied perpendicular 
to the CuO$_2$ planes \cite{Hill2001,Balci2002,Balci2003}. 
It was in this limit that a violation of the Wiedemann-Franz law was reported 
for \pcco (PCCO) \cite{Hill2001}, indicating that the low-temperature normal 
state could not be a Fermi liquid. Furthermore, differences in behavior 
between the electron-doped and hole-doped cuprate superconductors may also 
provide insights into the origin of their superconductivity.

A wealth of information about the normal and superconducting phases of
the hole-doped materials has been obtained from nuclear magnetic
resonance (NMR) experiments \cite{Asayama1996}. In NMR, the complex 
electron spin susceptibility $\chi(\vec{q},\omega)$ can be studied through the
spin-lattice relaxation rate \cite{Moriya1963} $\left[T_1^{-1}\propto
\int d\vec{q} |A_{\vec{q}}|^2 \chi''_{\perp}(\vec{q},\omega_n)/\omega_n
\right]$, the Knight-shift [related to $\chi'(0,0)$], and the spin-spin
couplings that probes $\chi'(\vec{q},0)$. Although $^{63}T_1T$ in the
normal state of PCCO is constant
\cite{Kumagai1990,Kumagai1990b,Bakharev1992,ZamboUnpublished}
as it would be for a Fermi liquid, it is enhanced by a factor $\approx
50$ above the independent-electron result \cite{Zheng2002}. 
Therefore complementing these results with information about 
spin-spin couplings can help to establish a consistent picture for 
the electron-doped cuprates as well.

Spin-spin couplings are often deduced from spin-echo decay rate
studies. In the standard Hahn echo sequence \cite{Hahn1950}, two
radio frequency (rf) pulses (hereafter referred 
to as $P1$ and $P2$) at the NMR frequency  are applied with a time 
separation of $\tau$, and a spin-echo signal forms centered at a time
$2\tau$ after the initial $P1$ pulse. With increasing $\tau$, the amplitude
of the echo signal decreases with a characteristic time $T_2$. It is caused by the
loss of spin-phase coherence, \textit{i.e.} irreversible dephasing 
that originates from one or more of the following mechanisms: 
1) direct nuclear spin-spin interaction (dipolar coupling); 
2) indirect (electron mediated) spin-spin interactions; 
3) spin-lattice relaxation; or 4) in general,
any kind of irreversible change in the local magnetic field (or electric
field gradient) experienced by the nuclei on the timescale of the spin-echo experiment
causes changes in the nuclear precession frequency and leads to irreversible
dephasing among the spins forming the echo.
Therefore the decay of the spin-echo height, $S(2\tau)$ can be decomposed into numerous
factors, each of which is often described by an exponential or a
Gaussian function. In the hole-doped systems the spin-lattice
fluctuations and the indirect spin-spin couplings are known to be the
relevant mechanisms; the later is enhanced by antiferromagnetic correlations.

\begin{figure}[!b]
\includegraphics[width=2.9in]{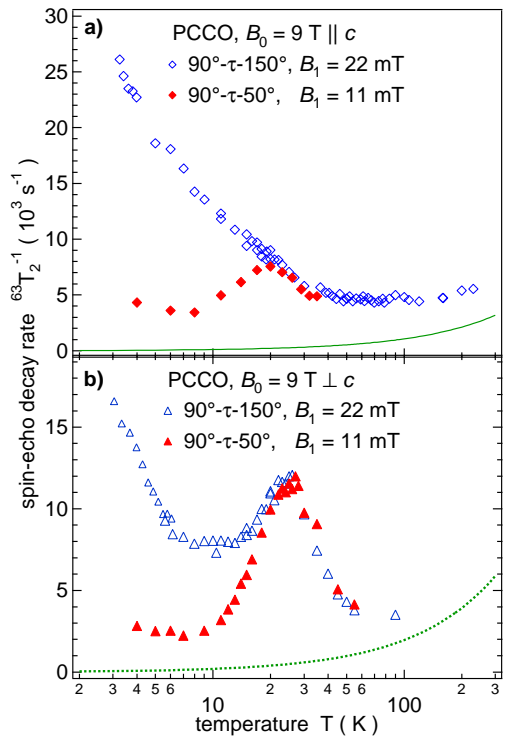}
\caption{Spin-echo decay rates $^{63}T_2^{-1}$ of PCCO obtained with
optimized pulses (solid symbols) and with a small angle refocusing
pulse at smaller $B_1$ (open symbols) as a function of temperature. The
solid (dashed) line shows the estimated Redfield contribution for
$B_0||c$ ($B_0$$\perp$$c$).}
\label{fig:T2highpower}
\end{figure}

In this Letter, we report measurements of $T_2$ in the normal
and superconducting phases of a single crystal of PCCO in a static
magnetic field $B_0=9$~T over the temperature range 2~K$<T<$200~K. 
For $T<25$~K, $T_2$ depends on the amplitude and duration
of the rf pulses used in the echo experiment. That is, irreversible 
dephasing of the spins involved in the echo formation results as a direct
consequence of the application of $P2$ in the two-pulse spin-echo sequence.
Although it is known that spin-echo decays in an inhomogeneously
broadened NMR line can depend on the pulses applied when nuclear 
spin-spin coupling is significant \cite{Klauder1962,Pennington1991,Pennington2001},
by adding a third pulse (hereafter referred to as $P_{nr}$) 
whose frequency differs from the NMR frequency we are able to rule
it out as the source for our observations. The results are interpreted as
evidence for the formation of an inhomogeneous electronic state that couples to
the rf pulses. At this time, we cannot state the nature of the inhomogeneous phase.

Single crystal PCCO samples were grown with a flux technique 
\cite{Peng1991,Brinkmann1996} and annealed in
argon at 900$^{\circ}$C for 48 hours. The doping
concentration is roughly optimum to maximize the
superconducting transition temperature at $T_{c0}\approx 22$~K, as
verified by zero-field-cooled magnetization measurements in 0.1~mT
magnetic field. The sample size used in this study was 3.5 mm x 2.5 mm x
35 $\mu$m. PCCO crystallizes in the \textit{T'}-tetragonal structure
leading to equidistant CuO$_2$ planes \cite{SaezPuche1983}, 
\textit{i.e.}\ all the Cu sites are equivalent and planar. 
(The planes are perpendicular to the $c$ axis
of the crystal.) From the diamagnetic effects of the sample on the
inductance of the NMR coil, we found that $T_c(B_0=9$~T$\perp c)\approx 15$~K. 
The silver coil was wound around a small piece of aluminium
powder in epoxy and the sample. The $^{27}$Al signal was used to 
calibrate the rf field $B_1$ at high temperatures.
Details of the Cu NMR spectra are to be reported elsewhere \cite{ZamboUnpublished}. 
They reveal that the origin of the inhomogeneous line broadening is magnetic,
and at both $B_0\parallel c$ and $B_0\perp c$ it is mainly the central 
transition that is measured.
The central transition has a full width at half maximum of $\Delta=19.5$~mT 
($B_0=9$~T, $T=2-150$~K).

In Fig.~\ref{fig:T2highpower} the open symbols show $T_2^{-1}$ obtained under the standard 
conditions for maximizing the spin-echo amplitude:  the rf field $B_1$ is large enough to 
cover fully the central transition of the $I=3/2$ nuclei, and both $P1$ and $P2$ has been 
adjusted to give the maximum echo height. Unlike hole-doped materials
\cite{Pennington1989,Pennington1991}, the observed decay is primarily
exponential for both $B_0\parallel c$ and $B_0\perp c$. At $T\approx 55$~K, both
$T_{2\parallel}^{-1}$ and $T_{2\perp}^{-1}$ depart 
from the very weak temperature dependence observed at higher temperatures. The increase
continues to the lowest temperature measured (2~K) for $B_0\parallel
c$. For the case $B\perp c$, there is a well-defined maximum in
$T_2^{-1}$ at $T\approx 25$~K. The solid symbols in Fig.~\ref{fig:T2highpower} 
show the results for the spin-echo decay rate when $B_1$ and the nutation angle of $P2$
are both considerably reduced. Clearly, the details of the pulses impact
significantly the values of $T_2$ at low temperatures.

\begin{figure}[!b]
\includegraphics[width=3.5in]{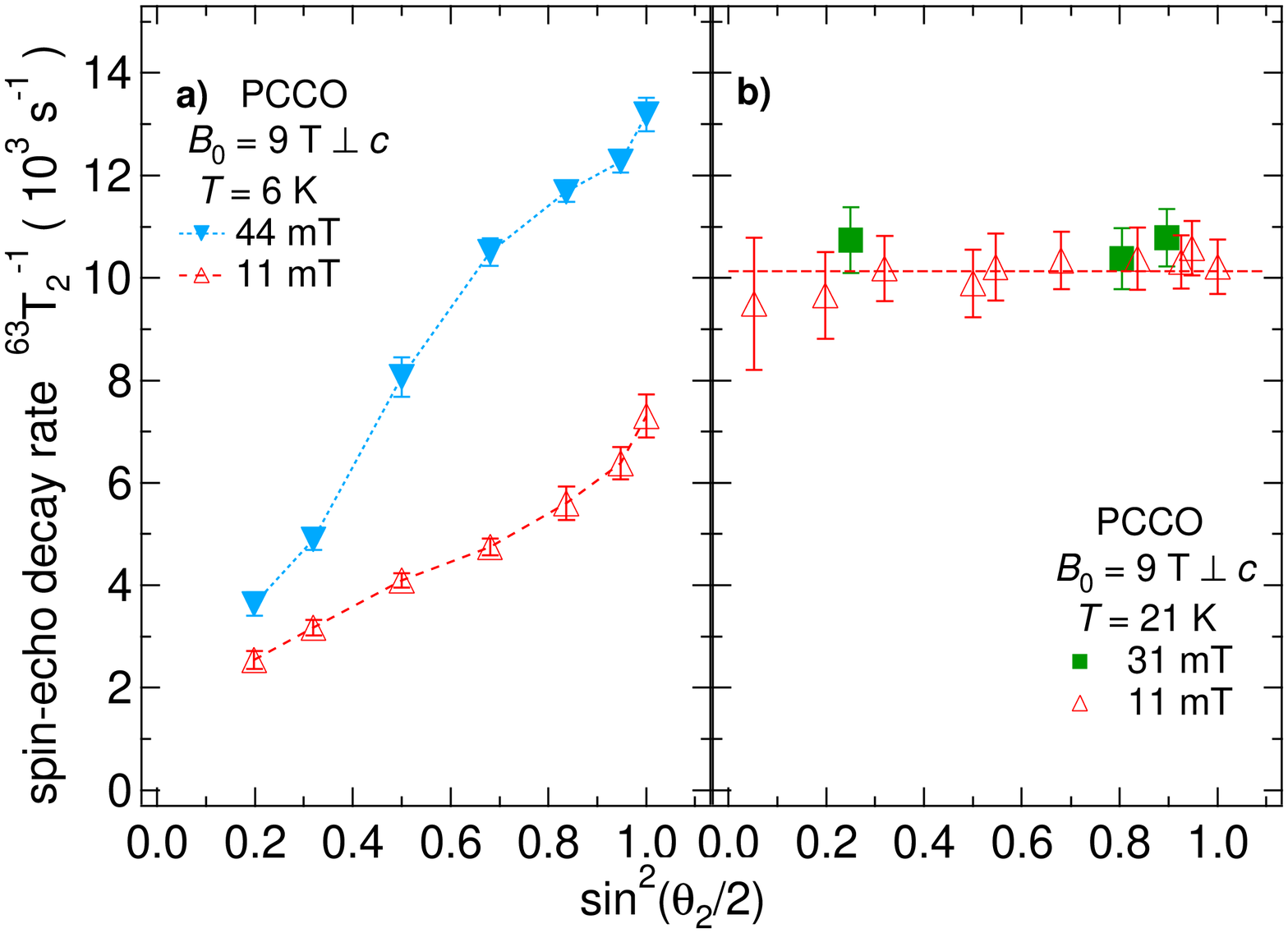}
\caption{Spin-echo decay rates $^{63}T_2^{-1}$ of PCCO as a function of
refocusing pulse angle ($\theta_2$) at various alternating magnetic fields $B_1$
in $B_0$=9~T$\perp$$c$ at \textbf{a)} $T=6$~K and \textbf{b)} $T=21$~K.
The dashed lines are guides to the eye. In order to avoid plotting a non-single 
valued function, only data points for $\theta_2<\pi$ are shown here. 
The values of $T_2^{-1}$ for $\theta_2>>\pi$ are close to the highest values
shown already on the figure.}
\label{fig:T2vsTheta2}
\end{figure}

Below, experiments are described that will justify our main conclusion
that $T_2$ is strongly affected by a local magnetic field distribution that can be modified by
the rf pulses.

\textit{Experiment I.)}
Fig.~\ref{fig:T2vsTheta2}a shows that the spin-echo decay rate
changes as the duration of $P2$, called $t_{w2}$ hereafter, is increased at
$T=6$~K for $B_1=44$~mT and 11~mT.
For reasons indicated below, the data are plotted as function of
$\sin^2\left(\theta_2/2\right)$, where $\theta_2=\alpha\gamma
B_1t_{2w}$, $\alpha=2$ for the central transition
of the $I=3/2$ nuclei \cite{Abragam1961}, and $\gamma$ is the gyromagnetic ratio. 
The insensitivity of the echo decay to pulse parameters at 
$T=21$~K is illustrated in Fig.~\ref{fig:T2vsTheta2}b.

\begin{figure}[!b]
\includegraphics[width=2.9in]{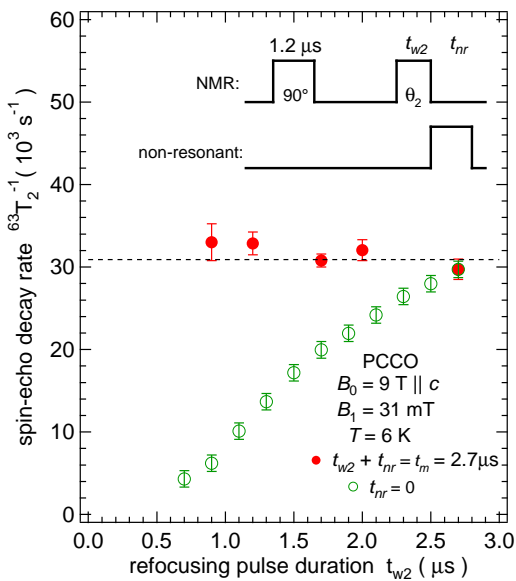}
\caption{Comparing spin-echo decay rates $^{63}T_2^{-1}$ obtained with
and without the application of an additional non-resonant rf pulse ($P_{nr}$), as discussed in
the text.}
\label{fig:T2woffrespulses}
\end{figure}

\textit{Experiment II.)}
The following two sets of measurements demonstrate that it is the
rf field itself that causes the dramatic change in $T_2$. The open
symbols in Fig.~\ref{fig:T2woffrespulses} show what happens if the duration of $P2$
in the standard NMR spin-echo experiment is varied. As before, the $T_2^{-1}$ increases 
as $t_{w2}$ increases up to the maximum value $t_m$. In the
second set of experiments a non-resonant $P_nr$ pulse [with a frequency 2~MHz
away from the center of the NMR spectrum (limited by the NMR tank circuit) and with
the same amplitude as the resonant one] is applied for a duration $t_{nr}$ just
after the $P2$ pulse. The combined length of $P_2$ and $P_{nr}$ is kept constant: 
$t_{w2}+t_{nr}=t_m$. It is evident (solid symbols in Fig.~\ref{fig:T2woffrespulses}) 
that the value of $T_2^{-1}$ depends only on the total duration $t_m$, and not on $t_{w2}$.
This result shows that the major change in the accumulated phase responsible for the reduction
of $T_2$ is caused by the rf pulse whether or not it is at the NMR frequency.

\begin{figure}[!t]
\includegraphics[width=2.9in]{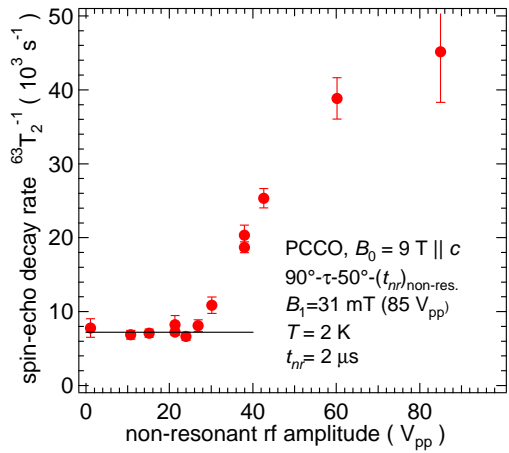}
\caption{Spin-echo decay rates $^{63}T_2^{-1}$ of PCCO obtained under
the influence of different non-resonant pulse amplitudes. All other
pulse parameters of $P1,P2,$ and $P3$ are the same.}
\label{fig:T2vsamplitude}
\end{figure}

Fig.~\ref{fig:T2vsamplitude} shows what happens at $T=2$~K if the
amplitude of $P_{nr}$ immediately following the $P2$ 
NMR pulse is varied. The details of the NMR pulses and the duration of
all the three pulses are kept constant. At this  low temperature, a
well-defined threshold or crossover rf field amplitude for the non-resonant pulse is
required to produce the additional reduction in $T_2$.

To discuss these results, the coupling Hamiltonian ($H'_i$) is written in
the following form \cite{mutualfootnote}
\begin{equation}
H'_i=I_{iz}\left[\sum_{j}a_{(i,j)z}I_{jz}+\hbar\gamma\delta B_i(t)\right],
\label{eq:coupling}
\end{equation}
where $i$ and $j$ refer to the $i^{\text{th}}$ and $j^{\text{th}}$ nuclear spins, $I_z$ is
the $z$-component of the nuclear spin operator, $a_{(i,j)z}$ is the internuclear coupling, and
$\delta B_i(t)$ is the \textit{deviation} of the local magnetic field at the $i^{\text{th}}$ nucleus as a
function of time ($t$) caused by processes \textit{other than nuclear spin-spin interactions}.
The first term of Eq.~\ref{eq:coupling} is often applied to the hole-doped cuprates, where
it leads to a Gaussian spin-echo decay with the time constant $T_{2G}$ \cite{Asayama1996}.
There is another term in the echo decay that is exponential in character,
called the Redfield contribution. It is uniquely determined by the
anisoptropic Cu spin-lattice relaxation rates \cite{Pennington1989,Pennington1991,Walstedt1995}. 
For PCCO, this contribution (shown in Fig.~\ref{fig:T2highpower}) is negligible at low $T$, as is
the contribution from direct nuclear couplings, which has the
upper bound of 470~s$^{-1}$. 
\ttg\ is a consequence of the indirect nuclear couplings that are enhanced by antiferromagnetic
fluctuations \cite{MMP1990}. A phenomenological, overdamped susceptibility peaked at the
antiferromagnetic wavevector was introduced \cite{MMP1990} as a way of modeling $T_1$ and $T_{2G}$.
Although we do not observe an unambiguous Gaussian component in our echo decays, it
does not rule out \textit{a priori} any influence of indirect spin-spin couplings
on the spin-echo experiments.
And given that the temperature dependence of $T_1^{-1}$ and $T_2^{-1}$
is opposite, it is likely that unrelated mechanisms govern the two
rates.

Now consider what is expected for the spin-echo decay when the first
term of Eq.~\ref{eq:coupling} dominates the behavior. Then,
pulse-induced spin flips of coupled neighbors alter the local magnetic
field with the application of $P2$, thus diminishing the
echo-refocusing. The echo decay is a function of
the probability that neighboring spins are flipped by the action of $P_2$
\cite{Pennington2001}, which 
can be written as $P(\theta_2)\propto\sin^2(\theta_2/2)$, as long as 
$B_1^{eff}=\alpha B_1>\Delta$ and $t_{w2}^{-1}<\Delta$, where $\Delta$ 
characterizes the linewidth. 
Therefore, it is expected that if internuclear coupling governs the spin echo decay,
the rate varies as a function of $P(\theta_2)$ only. The data of Fig.~\ref{fig:T2vsTheta2}
contradict this prediction, because the echo decay rates for the two
values of $B_1$ do not fall on the same curve. Furthermore, it is shown
in Fig.~\ref{fig:T2woffrespulses} that pulses which flip no spins ($P_{nr}$)
also control the decay rate. We conclude from these results that the
dependence of $T_2^{-1}$ on the parameters of $P_2$ occur independent of nuclear 
spin-spin interactions.

The observed effects arise from the second term in Eq.~\ref{eq:coupling}.
Consider a spatially varying magnetic field  that is changed by the rf field.
Now assume that the local field deviation at the $i^{\text{th}}$
nucleus after $P_1$ at $t=0$ is $\delta B_{i1}$, and it remains unchanged until 
the application of $P_2$, when it is changed to a value $\delta B_{i2}$. The
accumulated phase at the time of the echo formation ($t=2\tau$) is
$\Phi_i(2\tau)=\gamma\tau(\delta B_{i1}-\delta B_{i2})$,
and the shape of the echo decay for the ensemble of spins is proportional to
$\sum_i \cos\Phi_i(2\tau)$.
If the distribution of the $\delta B_{i1}-\delta B_{i2}$ happens to be Lorentzian, 
it can be shown that the echo decay is exponential.

Finally, we comment on the physical origin of $\delta B_i$. 
The rf pulses must reconfigure a spatial 
inhomogeneity to produce the observed phenomena. Such a
reconfiguration has been reported for rf-induced flux lattice annealing
\cite{Clark1999}. The inhomogeneities in PCCO are clearly different from the superconducting
state and they are not due to chemical inhomogeneities because the effects are
not only dynamic, but occur only at low temperatures. There are several candidate states 
discussed in the literature, including stripes or puddles
 \cite{Zaanen1989,Tranquada1995,Kivelson1998}, and
$d$-density waves \cite{Chakravarty2001}.
As long as these states are weakly pinned, an inhomogeneous local
magnetic field is produced. The formation of such a state at low
temperatures might be related to the dramatic changes
recently observed in tunneling spectroscopy \cite{Biswas2001,Biswas2002,Alff2003}.

In summary, $^{63}T_2$ measurements have been presented in the electron-doped high-$T_c$ 
superconductor PCCO over a broad range of temperatures and rf pulse
conditions. They show a substantial temperature dependence for $T_2$ and an unusual
dependence on the amplitude, duration and frequency of the rf pulses that is not
explained by the usual models applied to such materials. We propose that the
observed dependence of $T_2^{-1}$ on the rf pulse conditions indicates a
spatially varying local magnetic or electronic field of non-nuclear origin whose
configuration at low temperatures is changed by the rf pulses.  The
origin of this inhomogeneous electronic configuration is at present
undetermined.

\small ACKNOWLEDGEMENTS. The work was supported in part by the NSF
Grants DMR-0072524, DMR-0203806 and DMR-0102350. We also
acknowledge helpful discussions with N.~Curro, P.~C.~Hammel,
S.~Chakravarty, N.~P.~Armitage, M.~Horvatic, and C.~Berthier.


\begin{thebibliography}{30}
\bibitem[\citenamefont{Zamborszky}(2003)]{mutualfootnote}
\bibinfo{author}{} \bibnamefont{Generally, mutual spin-flips are
  included in Eq.~1. In practice, the spin-flips are suppressed when the
  relative shifts of the coupled nuclei differ by more than the internuclear
  coupling terms. The inhomogeneous broadening is usually taken to be large
  enough in the cuprates that this limit applies.}
\end{thebibliography}
\end{document}